\documentclass{article}

\begin{document}

\markboth{F. J. Oliveira}{Quantized Intrinsic Redshift in Cosmological General Relativity}

\title{Quantized Intrinsic Redshift \\ in Cosmological General Relativity}
\author{Firmin J. Oliveira $^\dagger$}

\maketitle

\begin{center}
{$^\dagger$ Joint Astronomy Centre, \label{fn:affil}
           Hilo, Hawai{`}i, U.S.A. \\
  Email: {firmin@jach.hawaii.edu} \\
  Address:  {660 N. A{`}ohoku Place, Hilo, Hawai{`}i, U.S.A. 96720, \\
            Telephone: 808-969-6539, \\
            Fax: 808-961-6516.}}
\end{center}

\begin{abstract}

There are now several analyses reporting quantized differences in the redshifts
between pairs of galaxies. In the simplest cases, these differential redshifts
are found to be harmonics of fundamental periods of approximately
$72 \, \, {\rm km \, s^{-1}}$ and  $37.5 \, \, {\rm km \, s^{-1}}\,$.
In this paper a wave equation is derived based on cosmological general relativity,
which is a space-velocity theory of the expanding Universe. The wave equation is
approximated to first order and comparisons are made between the quantized
solutions and the reported observations.
\end{abstract}

\section{Introduction}

There have been numerous, detailed observations and analyses of discrete differential
red shift found between pairs of galaxies, especially Tifft\cite{tifft-0}
and Napier \& Guthrie\cite{napier-1}. Attempts have been made in the past by
Cocke\cite{cocke-1} and Nieto\cite{nieto-1}
to develop a gravitational wave equation which could explain the results.

In this paper a linear second-order wave equation is derived based on the space-velocity
theory of cosmological general relativity (CGR) of Carmeli\cite{carmeli-0}. A spherically
symmetric metric is used with comoving coordinates. First order approximations are made
for small and large radial distances and the resulting equations are solved. The large
distance quantized solutions are compared to the observational data from differential
redshifts found between galaxies.

\section{Wave Equation}

The CGR spherically symmetric, comoving space-velocity metric (Ref. \cite{carmeli-2}, Eq. (A.5))
is defined by the line element
\begin{equation}
  ds^{2} = \tau^{2} dv^{2} - e^{\mu} dr^{2}
           - Q^{2} \left( d{\theta}^{2} + sin^{2}(\theta) d{\phi}^{2} \right) \, , \label{eq:sph-sym-metric}
\end{equation}
where coordinate $v$ is the radial velocity of expansion of the universe, and $\tau$ is the Hubble-Carmeli time
constant, its value is $\tau=12.486 \,{\rm Gyr}\,$ (Ref. \cite{carmeli-2}, Eq. (A.66), p. 138).
The functions $\mu$ and $Q$ are dependent only on the radial velocity $v$ and the radial coordinate $r$,
where $r, \theta$ and $\phi$ are the usual spatial spherical coordinates. 
  From Eq. \ref{eq:sph-sym-metric} the non-zero
elements of the metric $g_{\mu \nu}$ are
\begin{eqnarray}
  g_{0 0} & = & 1\, , \\ 
  g_{1 1} & = & -e^{\mu} \, , \\
  g_{2 2} & = & -Q^{2}\, , \\
  g_{3 3} & = & -Q^{2} sin^{2}(\theta) \, .
\end{eqnarray}

A postulate of this cosmological theory is that the metric $g_{\mu \nu}$ satisfies the Einstein field equations
\begin{equation}
  G_{\mu \nu}  =  R_{\mu \nu} - \frac{1}{2} g_{\mu \nu} R =  \kappa T_{\mu \nu}  \, , \label{eq:Einstein-eqn}
\end{equation}
where $ R_{\mu \nu}$ is the Ricci tensor, $R=g^{\alpha \beta} R_{\alpha \beta}$ is the scalar curvature, 
$T_{\mu \nu}$ is the momentum-energy tensor,
$\kappa = 8 \pi {\rm G} / (c^{2} \tau^{2})\,$, ${\rm G}$ is Newton's constant and $c$ is the speed of
light in vacuo. (Compare this to the Einstein equation in General Relativity theory where
$\kappa = 8 \pi {\rm G} / c^{4}$.) The momentum-energy tensor of a perfect fluid
 (Ref. \cite{carmeli-2}, Eq. (A.10)) is 
\begin{equation}
  T_{\mu \nu} = \rho_{eff}\, u_{\mu} u_{\nu}
               + p  \left( u_{\mu} u_{\nu} -  g_{\mu \nu} \right) \, ,
\end{equation}
where the effective mass density $\rho_{eff} = \rho - \rho_{c}$, where $\rho$ is the average mass density
of the Universe and $\rho_{c}$ is the  critical mass density, a {\em constant} in CGR given by 
 $\rho_{c} = 3/(8 \pi {\rm G} \tau^{2})$. Also, $p$ is the pressure, and $u^{\mu}$ is the four-velocity
\begin{equation}
u^{\alpha} = u_{\alpha} = \left(1, 0, 0, 0 \right). 
\end{equation}

The solution for Eq. \ref{eq:Einstein-eqn} was derived (Ref. \cite{carmeli-2}, Appendix A.4) with the results,
   \begin{eqnarray}
      Q & = & r\, , \label{eq:nu_solution} \\
  \nonumber \\
      e^{\mu} & = &  e^{\mu(r)} = \frac{1}{1 + f(r)} \, , \label{eq:mu_solution} \\
  \nonumber \\
      f(r) & = & \frac{1 - \Omega}{c^{2} \tau^{2}} r^{2} \, , \label{eq:fr} \\
  \nonumber \\
      p & = & \frac{c}{\tau} \frac{\left( 1 - \Omega \right)}{8 \pi {\rm G}} \, ,
   \end{eqnarray}
where the density parameter $\Omega = \rho / \rho_{c}$.

The linearized line element is defined by 
\begin{eqnarray}
  ds = d\left(\tau v \right) 
          - i e^{\mu(r)/2} dr
          - i r d\theta - i r \, sin(\theta) d\phi  \, ,  \label{eq:linear-line-elem-1}
\end{eqnarray}
where $i=\sqrt{-1}$.

 The linearized first order spinor differential equation follows by canonical correspondence\cite{ohara}
 from Eq. \ref{eq:linear-line-elem-1}  using spinors $\alpha_{\mu} \,$,
\begin{eqnarray}
  \frac{\partial{\Psi}}{\partial{s}}  =  
         \left\{ \alpha_{0} \frac{\partial{}} {\partial{\left( \tau v \right)}}
        +    \alpha_{1} \left( -i e^{-\mu(r)/2} \right) \frac{\partial{}}{\partial{r}}  \right. \nonumber \\
\nonumber \\
        +  \left. \alpha_{2} \left( \frac{-i}{r} \right) \frac{\partial{}}{\partial{\theta}}
        +         \alpha_{3} \left( \frac{-i}{r \, sin(\theta)} \right) \frac{\partial{}}{\partial{\phi}} \right\} \Psi
                 \, . \label{eq:linear-wave} 
\end{eqnarray}
Square Eq. \ref{eq:linear-wave} and assume the r.h.s. linearizes to the form of the d`Alembertian,
\begin{eqnarray}
 & &  \frac{\partial^{2}{\Psi}}{\partial s^{2}} =
  \left\{ \frac{\partial^{2}}{\partial{\left(\tau v\right)^{2}}}
        -e^{-\mu(r)} \frac{1}{r^{2}}\frac{\partial}{\partial{r}}
             \left( r^{2}\frac{\partial}{\partial{r}} \right) \right. \nonumber \\
\nonumber \\
  & & - \left. \frac{1}{r^{2}}
         \left[ \frac{1}{sin(\theta)} \frac{\partial}{\partial{\theta}}
                \left( sin(\theta) \frac{\partial}{\partial{\theta}} \right)
      + \frac{1}{sin^{2}(\theta)} \frac{\partial^{2}}{\partial{\phi^{2}}} \right] \right\} \Psi \, .  \label{eq:Psi_wave}
\end{eqnarray}

In CGR the condition for the expansion of the universe is defined by setting $ds=0$.
For the wave equation, it is assumed that the expansion of the universe corresponds to setting 
 $\partial^{2}{\Psi} / \partial{s^{2}} = 0$. With this condition,
Eq. \ref{eq:Psi_wave} becomes
\begin{eqnarray}
 \frac{1}{\tau^{2}} \frac{\partial^{2}{\Psi}}{\partial{v^{2}}} =
     e^{-\mu(r)} \frac{1}{r^{2}} \frac{\partial}{\partial{r}} \left( r^{2} \frac{\partial{\Psi}}{\partial{r}} \right)
                \nonumber \\
  \nonumber \\
        +\frac{1}{r^{2}}
  \left[ \frac{1}{sin(\theta)} \frac{\partial}{\partial{\theta}}
                \left( sin(\theta) \frac{\partial{\Psi}}{\partial{\theta}} \right)
                           + \frac{1}{sin^{2}(\theta)} \frac{\partial^{2}{\Psi}}{\partial{\phi^{2}}} \right]
                   \, .   \label{eq:stat-wave-sq3}
\end{eqnarray}

\section{Solving the Wave Equation}

We solve Eq. \ref{eq:stat-wave-sq3} by a separation of variables.
Assume that $\Psi(v,r,\theta,\phi) = \Psi_{0}(v)\, \Psi_{1}(r)\, \Psi_{2}(\theta)\, \Psi_{3}(\phi)$.
Substituting the composite function for $\Psi$ into Eq. \ref{eq:stat-wave-sq3} and dividing both sides by $\Psi$
we obtain
\begin{eqnarray}
& &   \frac{1}{\Psi_{0}(v)}\frac{d^{2}{\Psi_{0}(v)}}{\tau^{2} dv^{2}} =   \nonumber \\
\nonumber \\
& &   e^{-\mu(r)}
    \frac{1}{\Psi_{1}(r)} \frac{1}{ r^{2}} \frac{d}{dr} \left( r^{2} \frac{d\Psi_{1}(r)}{dr} \right)
       \nonumber \\        
  \nonumber \\
& &    + \frac{1}{r^{2}} \left[ \frac{1}{\Psi_{2}(\theta)} \frac{1}{sin(\theta)} \frac{d}{d\theta}
                \left( sin(\theta) \frac{d\Psi_{2}(\theta)}{d\theta} \right) \right. \nonumber \\
\nonumber \\
& &    +  \left. \frac{1}{sin^{2}(\theta)} \frac{1}{\Psi_{3}(\phi)}
                  \frac{d^{2}{\Psi_{3}(\phi)}}{d\phi^{2}} \right] \, . \label{eq:stat-wave-sq4} 
\end{eqnarray}
In Eq.\ref{eq:stat-wave-sq4}, since the left hand side is a funtion of $(v)$ only, while the
right hand side is a function of $(r,\theta,\phi)$ only, they must each equal a constant, $-D^{2}$.
Thus we can say
\begin{equation}
  \frac{1}{\Psi_{0}(v)} \frac{d^{2}{\Psi_{0}(v)}}{\tau^{2}dv^{2}} = -D^{2}  \, ,  \label{eq:V-wave-eq}
\end{equation}
\begin{eqnarray}
  -D^{2} = e^{-\mu(r)}
    \frac{1}{\Psi_{1}(r)} \frac{1}{ r^{2}} \frac{d}{dr} \left( r^{2} \frac{d\Psi_{1}(r)}{dr} \right)
       \nonumber \\
  \nonumber \\
     + \frac{1}{r^{2}}
   \left[ \frac{1}{\Psi_{2}(\theta)} \frac{1}{sin(\theta)} \frac{d}{d\theta}
                \left( sin(\theta) \frac{d\Psi_{2}(\theta)}{d\theta} \right)
             \right. \nonumber \\
\nonumber \\
     + \frac{1}{\Psi_{3}(\phi)} \frac{1}{sin^{2}(\theta)}
                \left.  \frac{d^{2}{\Psi_{3}(\phi)}}{d\phi^{2}} \right]  \, .  \label{eq:RTP-wave-eq}
\end{eqnarray}

Solutions for Eq. \ref{eq:V-wave-eq} have the form
\begin{equation}
  \Psi_{0}(v) = e^{\pm i D \tau v} \, ,
\end{equation}
where $D$ is called the {\em intrinsic curvature}.

Equation \ref{eq:RTP-wave-eq} can be split in two parts, one a function of $\,(\theta, \phi)$
only, the other a function of $\,(r)$ only, both equal the same constant $-L^{2}$. They simplify
to
\begin{eqnarray}
       \frac{1}{\Psi_{2}(\theta)} \frac{1}{sin(\theta)} \frac{d}{d\theta}
                \left( sin(\theta) \frac{d\Psi_{2}(\theta)}{d\theta} \right)  \nonumber \\
\nonumber \\
        + \frac{1}{sin^{2}(\theta)} \frac{1}{\Psi_{3}(\phi)} 
          \frac{d^{2}{\Psi_{3}(\phi)}}{d\phi^{2}} + L^{2} = 0 \, ,   \label{eq:TP-wave-eq}
\end{eqnarray}
and
\begin{eqnarray}
    & & \frac{1}{r^{2}} \frac{d}{dr} \left( r^{2}\frac{d\Psi_{1}(r)}{dr} \right) \nonumber \\
  \nonumber \\
    & &  +  D^{2} e^{\mu(r)} \Psi_{1}(r)
         - \frac{L^{2} e^{\mu(r)}}{r^{2}} \Psi_{1}(r) = 0 \, .  \label{eq:R-wave-eq3} 
\end{eqnarray}

The solutions for Eq. \ref{eq:TP-wave-eq} are the well known spherical harmonics
\begin{equation}
  \Psi_{2}(\theta) \Psi_{3}(\phi) =  P^{k}_{l}(cos(\theta)) e^{\pm i k \phi} \, , \label{eq:Theta_Phi_solution}
\end{equation}
where $P^{k}_{l}(cos(\theta)) $ are the associated Legendre functions and
\begin{eqnarray}
& &   k =  0, \pm 1, \pm 2, \pm 3, ... \, ,  \\
\nonumber \\
& &  L^{2} =  l \left( l + 1 \right) \, , \\
\nonumber \\
& &  l =  \mid k \mid, \; \mid k \mid + 1, \; \mid k \mid + 2,\, ... \,  .
\end{eqnarray}

To solve Eq. \ref{eq:R-wave-eq3} for $\Psi_{1}(r)$ we make first order approximations for $e^{\mu(r)}$.
Assuming $\mid \left( 1 - \Omega \right) r^{2} / \left( c^{2} \tau^{2}\right)  \mid \ll  1$
we have from Eqs.\, \ref{eq:mu_solution} and \ref{eq:fr}
\begin{equation}
  e^{\mu(r)} \approx  1 - \frac{\left(1 - \Omega \right) r^{2}}{c^{2} \tau^{2}} \, . \label{eq:e_mu_approx}
\end{equation}
Putting this approximation into Eq. \ref{eq:R-wave-eq3} and collecting terms we have
\begin{eqnarray}
  & &    \frac{1}{r^{2}} \frac{d}{dr} \left( r^{2}\frac{d\Psi_{1}(r)}{dr} \right)
      - \frac{ L^{2}}{r^{2}} \Psi_{1}(r) \nonumber  \\
  \nonumber \\
  & &  - \frac{\left( 1 - \Omega \right) } { c^{2} \tau^{2}}  D^{2} r^{2}\, \Psi_{1}(r) \nonumber \\
  \nonumber \\
  & &  +  \left[ D^{2}  + \frac{\left( 1 - \Omega \right) } {c^{2} \tau^{2}}  L^{2} \right] \Psi_{1}(r) 
       = 0.  \label{eq:R-wave-eq5}
\end{eqnarray}

\section{Approximation for Small ${\bf r}$, Quantization}

For small distance $r$ from a comoving source, such as a galaxy center,
and for small $L^{2}$, in Eq.  \ref{eq:R-wave-eq5}, drop the third term and
drop the second of the forth term, and subtitute  $L^{2} = l (l + 1)$ to obtain
\begin{eqnarray}
  \frac{1}{r^{2}} \frac{d}{dr} \left( r^{2}\frac{d\Psi_{1}(r)}{dr} \right)
         +  \left[ D^{2}
         -  \frac{l (l + 1)}{r^{2}} \right] \Psi_{1}(r) = 0 \, .  \label{eq:R-wave-eq-small-r}
\end{eqnarray}
Equation \ref{eq:R-wave-eq-small-r} is identical in form to that of Ref. \cite{nieto-1}, (Eq. 5.2). The
eigenfunctions are the two linearly independent spherical Bessel functions $j_{l}(r \, D)$ and $y_{l}(r \, D)$,
with eigenvalues determined at the boundary $r = c \tau$ where
 $j_{l}(c \tau D_{n l}) = 0$ and $y_{l}(c \tau D^{'}_{n l}) = 0$ for $n = 1, 2, 3, 4, ...$ and $l = 0, 1, 2, 3, ...$.
The eigenfunctions are related to the ordinary Bessel functions $J_{l}(x)$ by
\begin{eqnarray}
  j_{l}(x) & = & \sqrt{\frac{\pi}{2 x}} J_{l+1/2}(x)   \, \\
 \nonumber \\
  y_{l}(x) & = & (-1)^{l+1} \sqrt{\frac{\pi}{2 x}} J_{-l-1/2}(x) \, .
\end{eqnarray}

The first $6 x 6$ values of the zeros $c \tau D_{n l}$ of $j_{l}(x)$ are
\begin{eqnarray}
\begin{array}{c}
\begin{array}{ccccccc}
  n \; \backslash \;  l+1/2 & 1/2 & 3/2 & 5/2 & 7/2 & 9/2 & 11/2 \\
  ----  &  - & - & - & - & - & - \\
   1            &  3.14 & 4.49 & 5.76 &	6.99 & 	8.18  & 9.36 \\
   2            &  6.28 & 7.73 & 9.10 &	10.41 &	11.70 & 12.97 \\
   3            &  9.42 & 10.90 & 12.32 & 13.70 & 15.04 & 16.35 \\ 
   4            & 12.57 & 14.07 & 15.51 & 16.92 & 18.30 & 19.65 \\
   5            & 15.71 & 17.22 & 18.69 & 20.12 & 21.53 & 22.90 \\
   6            & 18.85 & 20.37 & 21.85 & 23.30 & 24.73 & 26.70 \\
\end{array}
\end{array}
\end{eqnarray}
For high order $n \ge N$, the zeros of $j_{l}(c \tau D_{n l})$ approach
\begin{eqnarray}
  \frac{{\rm lim}}{n \rightarrow \infty} \; c \tau D_{n l}  & = & \left(n + \frac{l}{2} \right) \pi \, , \\
  \nonumber \\
          n & = & N, N+1, N+2, N+3, ... \, , \\
          l & = & 0, 1, 2, 3, ... \, .
\end{eqnarray}

The discrete intrinsic curvature $D_{n l}$ are due to the gravitational
wave modes associated with the expansion of the Universe. It is assumed that the intrinsic
redshift $z_{i}$ observed in the wavelength of a photon is proportional to the intrinsic
curvature of the source of the photon,
\begin{equation}
  z_{i , n l} =  b \, c \, \tau D_{n l}  \, , \label{eq:zi_definition}
\end{equation}
where $b$ is a dimensionless positive constant. These eigen solutions are for small distances from some
comoving source. Notice that under the prevailing assumptions these solutions are independent of $\Omega$,
which means that they apply to all three phases of the expansion of the Universe: deceleration, constant
and acceleration\cite{behar-1}. A prediction of CGR is that the expansion is currently
accelerating\cite{carmeli-3}.

For $l=0$ or for large $n$,  Eq. \ref{eq:zi_definition} does yield a linear relationship of
intrinsic redshift to intrinsic curvature.  However, these are for small distances, whereas in the reports
to be discussed later on, the galaxies can be separated by large distances of a half megaparsec to 
a few tens of megaparsec.  We suggest that it may be applicable in the analysis of the redshift of quasars
presumed to be in near proximity to a host galaxy as reported in Ref. \cite{karlsson-2},
 \cite{chu-1} and \cite{galianni-1}.

\section{Approximation for Large ${\bf r}$}

For large $r$ make the approximation
\begin{equation}
  \frac{1}{r^{2}} \frac{d}{dr} \left( r^{2}\frac{d\Psi_{1}(r)}{dr} \right) \approx 
               \frac{d^{2} \Psi_{1}(r)}{dr^{2}} \, .
\end{equation}
For large distance $r$ from a comoving source and for small $L^{2}$, in Eq. \ref{eq:R-wave-eq5},
ignore the second term and the second of the fourth term to obtain
\begin{eqnarray}
      \frac{d^{2} \Psi_{1}(r)}{dr^{2}} 
      + \left[ D^{2} - \frac{\left( 1 - \Omega \right) } { c^{2} \tau^{2} } D^{2} r^{2} \right]
                \Psi_{1}(r)  = 0. \label{eq:R-wave-eq-large-r}     
\end{eqnarray}
There are three possible solutions to consider, depending on whether $\Omega > 1$, $\Omega = 1$ or
$0 \le \Omega  < 1$.

\subsection{$\Omega > 1$, Decelerated Expansion}

For $\Omega > 1$, generalized power series solutions are required for
 Eq. \ref{eq:R-wave-eq-large-r}. A solution for $\Psi_{1}(u)$ has the form
\begin{eqnarray}
  \Psi_{1}(u) & = & \sum^{\infty}_{k=0}{a_{k} u^{k}} \, ,  \label{eq:Psi_series} \\
  \nonumber \\
   u & = & \sqrt{\alpha}\, r \, ,
\end{eqnarray}
where $\alpha = \sqrt{\Omega-1} D / \left( c \tau \right)$.
When $\Psi_{1}(u)$ is substituted into Eq. \ref{eq:R-wave-eq-large-r} the values for the coefficients
are
\begin{eqnarray}
  a_{2} & = & -\frac{\gamma \, a_{0}}{2}  \, , \\
  \nonumber \\
  a_{3} & = & -\frac{\gamma \, a_{1}}{6}  \, , \\
  \nonumber \\
  a_{k + 2} & = & \frac{-\left( \gamma \, a_{k} +  a_{k - 2}\right)}
           {\left( k + 2 \right) \left( k + 1 \right)}  \,  , \\
  \nonumber \\
  k & = & 2, 3, 4, ... \, ,
\end{eqnarray}
where $\gamma = D c \tau / \sqrt{\Omega - 1}$.  We can let $a_{0}=1$ and $a_{1} = 0$. Thus all the odd
numbered coefficients $a_{2 k + 1}$ vanish.  The condition for the series Eq. \ref{eq:Psi_series}
to converge absolutely by the ratio test is
\begin{eqnarray}
  \lim_{k \rightarrow \infty} \mid \left( \frac{a_{2(k + 1)} u^{2}}{a_{2k}} \right) \mid = \rho \, ,
\end{eqnarray}
where $\rho < 1$. There is apparently no need for any discrete values of the parameter $D$.

\subsection{$\Omega = 1$, Constant Expansion}

For  $\Omega = 1$, Eq. \ref{eq:R-wave-eq-large-r} has the solutions
\begin{equation}
  \Psi_{1}(r) = \Psi_{c} e^{\pm i D r}  \, . \label{eq:R-wave-eq-large-r-omega-gt1}
\end{equation}
These solutions allow continuous values for $D$.

\subsection{$0 \le \Omega < 1$, Accelerated Expansion, Quantization}

For $0 \le \Omega < 1$, Eq. \ref{eq:R-wave-eq-large-r} is the equation of the quantum harmonic oscillator
with solutions involving the Hermite polynomials $H_{n}(\chi)$ of order $n$,
\begin{equation}
  \Psi_{1}(\chi) = \Psi_{c} \left( e^{-\chi^{2}/2} \right) H_{n}(\chi) \, ,
\end{equation}
where $\chi = \left( 1 - \Omega \right)^{1/4} \left( D / c \tau \right)^{1/2} r$, and $\Psi_{c}$ is a
constant. The solutions have the requirement that
\begin{eqnarray}
  \frac{c \tau D_{n}}{\sqrt{1 - \Omega}} = 2 n + 1 \, , \label{eq:R-wave-harm-quantized}
\end{eqnarray}
where $n = 0, 1, 2, 3, ...$.

As stated above, the intrinsic redshift $z_{i}$ is assumed to be proportional to the intrinsic
curvature $D_{n}$,
\begin{eqnarray}
  z_{i , n} = b \, c \, \tau \, D_{n} = Z_{0} \, \left( 2 n + 1 \right) \, ,
      \label{eq:zi_n_large_r} \\
\nonumber \\
Z_{0} = b \sqrt{1 - \Omega} \, , \label{eq:Z0}
\end{eqnarray}
where $b$ is a dimensionless positive constant.

\subsection{Theory Versus Experiment}

  We will now discuss the quantized version of the solutions for large $r$ given by
  Eq. \ref{eq:zi_n_large_r}. This equation is valid during the accelerated expansion stage.
  The idea put forth here is that the wave equation governed the
  accelerated expansion of the Universe and created states of quantized intrinsic curvature in
  the fluid. When the fluid coalesced to form gas clouds, the quantized curvature
  states persisted.

  Assume that $b = 1$. From Eq. \ref{eq:Z0} we get
  \begin{equation}
    Z_{0}^{2} = 1 - \Omega  \, .  \label{eq:Omega_less_one}
  \end{equation}
  A combined redshift formula\cite{karlsson-2} will show the relationship of the intrinsic
  redshift to the total redshift $z_{t}$,
  \begin{equation}
    1 + z_{t} =  \left( 1 + z_{c} \right) \left( 1 + z_{r} \right) \left( 1 + z_{i} \right) \, ,
         \label{eq:z_{total}}
  \end{equation}
  where $z_{c}$ is the cosmological redshift due to the  Hubble expansion and $z_{r}$ is a random
  component due to all other effects. From Eq. \ref{eq:z_{total}}, the difference of total redshift
  between a galaxy in intrinsic state $n_{2}$
  and another galaxy in intrinsic state $n_{1}$, where $n_{1} < n_{2}$,  is 
  \begin{eqnarray} 
   z_{t,2} - z_{t,1} 
      = & & \left( 1 + z_{c,2} \right) \left( 1 + z_{r,2} \right) \left( 1 + z_{i,n_{2}} \right)
            \nonumber \\
  \nonumber \\
        & & - \left( 1 + z_{c,1} \right) \left( 1 + z_{r,1} \right) \left( 1 + z_{i,n_{1}} \right) \, .
            \label{eq:DeltaZ_total}
  \end{eqnarray}
  Assume that the cosmological components $z_{c, m}$ are small and approximately equal for both
  galaxies and that their random components are negligible. Combining Eqs. \ref{eq:zi_n_large_r} and
  \ref{eq:DeltaZ_total} and retaining only the first order terms  we get differential redshift
  \begin{eqnarray}
    \Delta{Z}_{k}  = z_{t , 2} - z_{t, 1} \approx  z_{i , n_{2}} - z_{i , n_{1}} =  2 \, Z_{0} \, k \,
            , \label{eq:DeltaZ}
  \end{eqnarray}
  where $k = n_{2} - n_{1}$. From Eq. \ref{eq:DeltaZ}  define the differential intrinsic velocity 
  \begin{equation}
    \Delta{V}_{k} = c \Delta{Z}_{k} = 2 \, c \, Z_{0} \, k   \, . \label{eq:DeltaV}
  \end{equation}

  Now, we will briefly discuss two reports on the quantization observed in the differential
  redshifts from normal galaxies.
  In Ref. \cite{tifft-0}, $31$ pairs of galaxies were observed in radio and some in optical.
  Their redshift differentials were found to have a fundamental periodic velocity of
  $\Delta{V} = 72 \, {\rm km} \, {\rm s}^{-1}$. We use $c=299792.456 \, {\rm km} \, {\rm s}^{-1}$
  for the vacuum speed of light. To reproduce these periodicities, set 
  $Z_{0} = 36 / c = 1.201 \times 10^{-4}$.  From Eq. \ref{eq:Omega_less_one}, this would correspond to
  a density parameter $1 - \Omega = 1.442 \times 10^{-8}$. From Eq. \ref{eq:DeltaV} we have
  \begin{eqnarray}
    \Delta{V}_{k} & = & 2 \, c \, Z_{0} \, k = 72 \, k \, \,{\rm km} \, {\rm s}^{-1}  ,
           \label{eq:DeltaV_72} \\
          k & = & 1, 2, 3, 4 \, .
  \end{eqnarray}
  Equation \ref{eq:DeltaV_72}  gives the sequence of periodic velocities
  $72,\, 144,\, 216\,$ and $288\,$ $\left[{\rm km} \, {\rm s}^{-1}\right]$
  that appear as peaks in (Fig. 1) of Ref. \cite{tifft-0}.

  In Ref. \cite{napier-1}, $97$ galaxies were observed,  where the velocity of the galaxies extended out to
  $2600 \, {\rm km} \, {\rm s}^{-1}$. Their redshift differentials were determined, relative to the Galactic center, 
  with a fundamental periodic velocity of $\Delta{V} = 37.5 \, {\rm km} \, {\rm s}^{-1}$ obtained with
  amazing regularity.  Refer to their (Fig. 4) of that report.  To reproduce these periodicities, set 
  $Z_{0} = 18.75 / c = 6.2543 \times 10^{-5}$.  This would correspond to a density parameter
  $1 - \Omega = 3.9116 \times 10^{-9}$. From Eq. \ref{eq:DeltaV} we have
  \begin{eqnarray}
    \Delta{V}_{k} & = & 2 \, c \, Z_{0} \, k = 37.5 \, k \, \,{\rm km} \, {\rm s}^{-1} , \label{eq:DeltaV_37.5} \\
             k & = & 1, 2, 3, ..., 38 \, .
  \end{eqnarray}
  It is evident that Eq. \ref{eq:DeltaV_37.5} gives the sequence of periodic velocities that appear as peaks
  in (Fig. 4) of Ref. \cite{napier-1}, from $37.5 \, {\rm km} \, {\rm s}^{-1}$ up to
  $1425 \, {\rm km} \, {\rm s}^{-1}$.

  In closing we note that these are only preliminary analyses to show the efficacy of quantized intrinsic redshift.
  A more thorough analysis would study galaxy populations in terms of cosmological redshift and parametric density.

  The author is grateful to the Joint Astronomy Centre, Hilo, Hawai{`}i for many years of employment and support,
  and to Professor Moshe Carmeli for his theories and communications.

\end{document}